# Attosecond Electron Microscopy

Dandan Hui†, Husain Alqattan†, Mohamed Sennary, Nikolay V. Golubev and Mohammed Th. Hassan*.

Department of Physics, University of Arizona, Tucson, AZ 85721, USA

*Corresponding author. Email: mohammedhassan@arizona.edu
†These authors contributed equally to this work.

**Abstract:**

The electron motion in atoms and molecules is at the heart of all phenomena in nature that occur outside the nucleus. Recently, ultrafast electron and X-ray imaging tools have been developed to image the ultrafast dynamics of matter in real time and space. The cutting-edge temporal resolution of these imaging tools is on the order of a few tens to a hundred femtoseconds, limiting imaging to the atomic dynamics, hence the electron motion imaging remains beyond the reach. Here, we achieved attosecond electron imaging temporal resolution in a transmission electron microscope—orders of magnitude faster than the highest reported imaging resolution—to demonstrate, which we coin it as "attomicroscopy", to image the field-induced electron dynamics in neutral multilayer graphene. Our results show that the electron motion between the carbon atoms in graphene is due to the field driven electron dynamics in the conduction band and depends on the field waveform, strength, and polarization direction. This attomicroscopy imaging provides more insights into the electron motion of neutral matter in real time and space and would have long-anticipated real-life attosecond science applications in quantum physics, chemistry, and biology.

**Main Text:**

In the last two decades, the generation of attosecond Extreme Ultraviolet (XUV) pulses—based on strong field interactions and high-harmonic-generation (HHG) processes—has provided attosecond temporal resolution in spectroscopy measurements that are required to freeze and trace electron motion on its native time scale (*1-8*). However, the high photon energy of the XUV pulses limits the study of electron dynamics to ionized systems in strong-field interactions and restrains the investigation of electron dynamics in neutral systems. Moreover, femtosecond and attosecond spectroscopy provide crucial information on the evolution of atomic and electron motions in the time domain, although these motion trajectories in spatial dimensions remain beyond reach. Hence, ultrafast electron microscopy (UEM), diffraction (UED), and X-ray imaging tools were developed for ultrafast imaging dynamics in real time and space (*9-11*). Despite the remarkable advances that have been made in electron imaging techniques, the imaging time response remains limited to a hundred femtoseconds, leaving electron motion imaging out of reach (*12-34*). Recently, the generation of attosecond electron pulses in a train form inside the microscope—using laser pulses of several tens or hundreds of femtoseconds—has been reported (*31-33, 35, 36*). However, such a train of electron pulses can't be used in practical time-resolved electron imaging measurement since the dynamics is probed periodically every half-cycle. In this case, the imaging resolution is defined by the temporal profile of the entire envelope of the train pulses. Hence, obtaining the required attosecond temporal resolution to image electron motion in matter remains unfeasible.

In this work, we generated a single attosecond electron pulse by optical gating (*18, 37*) and controlled the free-electron pulses in the subfemtosecond time window with a polarization-gated half-cycle laser pulse (*38*). Accordingly, we attain and demonstrate the attosecond temporal resolution in electron microscopy imaging and establish the attomicroscopy. We utilized attomicroscopy to image the electron motion dynamics in neutral multilayer graphene by means of attosecond electron diffraction. The attomicroscopy imaging tool introduced in this work enables the study of the attosecond electron dynamics of matter in real time and space domains and opens a new avenue for real-life applications of attosecond science. Moreover, this long-awaited imaging of electron motion in action can reveal electron dynamics in complex and quantum systems, and promises to break new ground in physics, chemistry, biochemistry, and biology (*19*).

**Attomicroscopy: generation of a single attosecond electron pulse**

Attomicroscopy (setup is shown in Fig. 1A) is established to generate a single attosecond electron pulse (AEP) inside the electron microscope by the optical gating approach (Fig. 1B) (*18, 19, 37*). The basic principle of the electron gating is shown in Fig. 1C. Initially, ultrafast free-electron pulses are generated inside the microscope by a photoemission process using ultraviolet (UV) laser pulses. Then, few-cycle laser pulses (carried at a central wavelength of 750 nm, with a passively stabilized carrier-envelope phase) are divided into two beams, as shown in Fig. 1D. The first beam (pump pulse) is used to trigger the electron dynamics of the system under study. The second beam (gating pulse) goes through a polarization gating (PG) process (illustrated in Fig. 1E, and explained in Materials and Methods in the Supplementary Materials (SM), section I&II)— which has been utilized previously to generate single isolated attosecond XUV pulses (*38, 39*)—to produce a polarization-gated laser pulse. The output pulse from PG setup henceforth called the optical gating pulse (OGP) has a linear polarized half-cycle field in the middle and circular polarized field components at the sides (as shown in the shaded red area in Fig. 1E). The linear half-cycle in OGP has an estimated full width at half maximum (FWHM) of 625 attosecond (as). Then, the OGP is used to optically gate the free electron pulses inside the microscope. The optical gating process (*18, 19, 37*) (illustrated in Fig. 1B) can be briefly summarized as follows: the OGP interacts with the main electron pulses by inelastic scattering on a transmission electron microscope (TEM) dense-mesh aluminum grid placed on top of the sample (Fig. 1B). Note that aluminum was chosen as a gating medium due to its broad frequency response (*40*). The free electrons couple efficiently with the linearly polarized part of the field (the interaction with the circularly polarized sides of the field is minimal (*41*)) and exchange momentum. Hence, these electrons are gated in a temporal window similar to the duration of the linearly polarized half-cycle of the OGP, which is on the order of the sub-fs (~625 as) timescale. Hence, a single AEP is generated (per each OGP) directly on the top of the sample, maintaining its attosecond temporal profile in imaging experiments (see SM section II), and enhance the temporal resolution of the TEM to the subfemtosecond time scale. The number of gated electrons in the AEP is estimated to be 0.1% of the total number of electrons (~$10^6$ electrons/second) generated from the photocathode, which is sufficient to probe and image the attosecond electron motion dynamics as we report next.

**Imaging quantum electron motion in graphene**

The photoinduced electric current and the ultrafast carrier dynamics in graphene were studied earlier (*42-44*). These dynamics are strongly correlated to the structure in the space domain. So, the access to the electron motion dynamics in both real time and space is crucial and would provide more insights into the underlying physics of the light field driven dynamics in graphene and open the door to engineering graphene-based ultrafast optoelectronics devices.

In a strong light field, there are two induced electron dynamics in graphene (Fig. 2A): (i) The interband dynamics caused by the population transfer between valance (VB) and conduction (CB) bands and the related coherence dynamics, and (ii) The intraband dynamics generated due to the electron motion within the CB (*42*). These field driven sub-half-cycle electron dynamics can be traced in real time and space by recording electron diffraction snapshots of the multilayer graphene. Hence, as a first demonstration of attomicroscopy, we conducted a time-resolved attosecond electron diffraction imaging experiment to study the strong field-induced electron motion dynamics in a single-crystal multilayer (~6 layers) graphene (sample preparation and characterization are explained in the SM, section III). The generated AEPs is used to probe the electron motion dynamics of a graphene, which is induced by a 5-fs pump laser pulse.

In our experiment, the field strength of the driver field is estimated to be ~2.5 V/nm, and the OGP is set at a low power value (15 mw) to ensure that it is not triggering the graphene electron dynamics. Then, we recorded the diffraction pattern of the graphene (Fig. 2B) as a function of the time delay between the OGP and the pump pulses with a step size of 300 as. The retrieved time-resolved attosecond electron diffraction results—after subtracting the dark diffraction pattern (at $\tau = -1$ fs), as explained in SM, section IV—show that the scattering intensity of the Bragg diffraction peaks is modulating as a function of time in the presence of the driver field. The average intensity changes of the 1st-, 2nd-, and 3rd-order scattered peaks (highlighted by the white, yellow, and red lines and circles in Fig. 2B, respectively) are shown in Fig. 2C-E, respectively. The scattering intensity of the three diffraction order peaks are oscillating similarly and with the same frequency of the pump field. Remarkably, the ability to resolve the half-cycle scattering intensity modulations (Fig. 2C-E) demonstrates the attosecond electron imaging resolution of the attomicroscopy. Worth notes, we performed a few additional experiments to confirm the achieved resolution. In the first confirmation experiment, after recording the measurements presented in Fig. 2c-e, we blocked the OGP and repeated the electron diffraction measurements. The measured diffraction patterns showed no intensity change in the Bragg spots (at any order) in the time window between -1 and 25 fs, as expected. Similarly, the time-resolved diffraction measurements showed no modulation when we reduced the pump field strength to <2 V/nm (low-field regime)(*42*). Remarkably, we observed the scattering intensity modulation signal again by ramping up the power of the pump pulse back to the high field strength of 2.5 V/nm. These results show that the measured diffraction modulation is solely related to the graphene dynamics. Finally, repeating the same measurements under the same conditions but with gating medium (aluminum grid) removed. The results show no resolved oscillations confirming the measured diffraction results in Fig. 2c-e are occurred due to the generation of the subfemtosecond gated electron pulse at the aluminum grid. Hence, all these confirmation measurements verify the attained attosecond resolution.

Furthermore, the measured diffraction peak oscillations in Fig. 2C-E carries the signature of the field-induced electron motion in real space. To understand the measured electron dynamics and the underlying physics, we performed fully quantum-mechanical simulations of the laser-driven electron motion dynamics in multilayer graphene using the nearest-neighbor tight-binding model (*45*). The detailed description of the simulation and calculations are explained in the SM (Sections V). In our model, we calculated the evolution of the electron density in reciprocal space by solving the Bloch equation (*46, 47*):

$$i\hbar \frac{\partial}{\partial t} \rho_{m,n}(k,t) = \left(E_m(k_t) - E_n(k_t)\right)\rho_{m,n}(k,t)$$
$$+ E(t) \cdot \{D(k_t), \rho(k,t)\}_{m,n} - i\frac{1-\delta_{m,n}}{T_d}\rho_{m,n}(k,t),$$

(1)

where $E(t)$ is the applied electric field, $\rho_{m,n}(k,t)$ denotes the matrix element of the density matrix $\rho(k,t)$, $E_i(k_t)$ are the energies of the bands, $D(k_t)$ is the matrix of transition dipole moments, the commutator symbol "{}" is defined as $\{A,B\} = AB - BA$, and $T_d$ is the interband dephasing time. The simulations are performed in the time-dependent crystal momentum frame, which evolves according to the Bloch acceleration theorem:

$$k_t = k + {e}/{\hbar} A(t),  \quad (2)$$

where $k$ is the wave vector of the electron before the interaction with the applied vector potential $A(t) = -\int_{-\infty}^{t} dt' E(t')$. Notably, to have a solid correlation between the simulation and experiment results, we utilized in the presented calculation the driver field (Fig. S2 in SM) used in the diffraction experiment, which is measured by the all-optical light field sampling methodology (*7, 8, 48*).

The computed density matrix $\rho(k, t)$ provides an access to the observable properties of graphene such as the electron current $J(t)$ and the real space charge density $Q(r, t)$ distribution, which are calculated as explained in SM (Sections III and VIII). Importantly, both these quantities can be decomposed in two components: non-coherent and coherent contributions which, in turn, reflect the intra- and interband electron dynamics induced by the pump field.

Accordingly, we calculated the intra- and interband currents and fitted them to the measured diffraction intensity modulations in Fig. 2C-E, as explained in SM (Section VI). Remarkably, the fitted curve (plotted in red line in Fig. 2C-E) are in a good agreement with the main diffraction intensity oscillations of the 1st-, 2nd-, and 3rd-order peaks. The small deviation on the sides may be attributed to the underestimation of the dispersion effect experienced by the pump field with respect to the sample position in the performed field sampling measurement. The fitting results show that the contribution of the intra-band current (presented by fitting parameter a) is significantly higher respect to the contribution of the intra-band current (fitting parameter b). This ratio is estimated to be ~87%, ~85%, and ~80% from the fitting of the first, second, and third diffraction peaks' dynamics, respectively. These results indicate that the diffraction intensity oscillations are mainly due to the intraband current (electron motion within the CB).

Moreover, we conducted the attosecond diffraction measurements at different field strengths (SM, Section VII). The results show that the average diffraction intensity oscillations start to show up at certain threshold (between 1.8 and 2 V/nm)—which agrees with the previously reported photoinduced current measurements from graphene (*42*)—and increase linearly as a function of field strength as shown in Fig. S5. This linear tendency indicates that the measured diffraction oscillations are attributed to the intraband dynamics—supporting the conclusion of the fitting results—since the interband and the coherence dynamics are nonlinearly sensitive to the driver field.

The experiment results shown in Fig. 2C-E drive the direction of the conducted simulation study to study the electron motion in reciprocal and real space. First, to reveal the electron dynamics in the reciprocal space, we utilized our quantum-mechanical model (Eq. ( 1 )) to calculate the total population dynamics of the CB and plot it in Fig. 3A (red line), in contrast with the vector potential of the pump field (blue lines). The CB population increases continuously until reaches its maximum. After the driver field, the remaining excited electrons in the CB take a long time—on the order of a few tens to hundreds of femtoseconds (outside the time window of this study)—to relax back to the ground state. Furthermore, we calculated the electron density distribution (EDD) dynamics in the reciprocal space as it evolved over time (as shown in Fig. 3B). Initially, at the arrival of the field ($\tau = 1$ fs), the excited electrons are localized around the Dirac point as illustrated in Fig. 3B (I). Then, as the field strength increases at $\tau = 5.6$ fs, the electrons migrate to the positive $K_x$ direction, as depicted in Fig. 3B (II). At $\tau = 6.8$ fs, the electrons are return back towards the $K_x$ negative direction following the field shape, as shown in Fig. 3B (III). At the field highest half-cycle strength ($\tau = 8$ fs), the electrons move in the direction of a higher positive value of $K_x$ (shown in Fig. 3B (IV)) than the electrons at the maximum strength of the first half-cycle (Fig. 3B (II)), indicating that the electron displacement depends on the strength of the driver field. Moreover, in Fig. 3B (IV), the number of excited electrons at the reciprocal space increases and the EDD spreads in positive and negative directions of $K_y$. At $\tau = 10.6$ fs, the EDD is minimal around the Dirac point and more spread in both directions (positive and negative values) of $K_x$ and $K_y$, as shown in Fig. 3B (V). After the end of the field (at $\tau = 18$ fs); the EDD is all over the reciprocal space (Fig. 3B (VI)). The full mapping of the electron dynamics in the reciprocal space, provided in SM Video 1.

Second, from our quantum-mechanical simulation we obtained the electron motion dynamics around the graphene carbon atoms in real space. Figure 4A shows snapshots of the EDD at different time instants (a full movie is provided in SM Video 2). Note, the carbon atoms of the top graphene layer and the bonds between these atoms are shown in black solid circles and lines, respectively. Before the field arrival, the electrons are in the equilibrium state, and no EDD change occurs around the graphene carbon atoms. At the maxima of the first half-cycle ($\tau = 5.6$ fs), the change in the EDD is observable as shown in Fig. 4A (I). Hence, the EDD is more localized around A, B, & C than $\bar{A}$, $\bar{B}$, & $\bar{C}$ atoms. Also, the EDD intensity between $B\bar{B}$, $C\bar{C}$, and $\bar{A}A$ (from the next unit cell) atoms are lower than between $A\bar{B}$, $A\bar{C}$, $B\bar{A}$ and $C\bar{A}$ atoms in space. When the field flips the direction at $\tau = 6.8$ fs (Fig. 4A (II)), the electrons migrate towards the $\bar{A}$, $\bar{B}$, & $\bar{C}$ atoms and the EDD localization differences between A, B, & C and $\bar{A}$, $\bar{B}$, & $\bar{C}$ atoms decrease. The EDDs between all the atoms are almost similar. Furthermore, near the field maxima ($\tau = 8.0$ fs), the EDD localization difference around and between the atoms is maximized as illustrated in Fig. 4A (III). After a half-cycle (at $\tau = 9.4$ fs), the field flips the direction and the EDD difference around the atoms becomes less as shown in Fig. 4A (IV). At $\tau = 10.6$ fs (the maxima of the third half-cycle of the field), the localizing difference of EDD increase again, as illustrated in Fig. 4A (V), due to the change in field direction from Fig. 4A (IV). Although the EDD difference in Fig. 4A (IV) remains less than Fig. 4A (III) and higher than in Fig. 4A (I) because in this case the field strength is less than the main half cycle at $\tau = 8.0$ fs but higher than the field strength of the first half cycle at $\tau = 5.4$ fs. This indicating that the EDD in real space depends on the direction and the strength of the driver field similar to the electron motion in the CB (Fig. 3B).

Furthermore, the capability to decompose the calculated intra- and interband dynamics in our simulations allowed us to decouple their contributions to the electron motion dynamics of graphene in the real space (see SM section V). Hence, we calculated snapshots of the EDD considering solely the effect of the intraband current and plot it in Fig. 4B. These results show that the intraband dynamics is causes the EDD displacement between the carbon atoms in real space, which explains the scattered electron intensity oscillations observed in our diffraction measurements in Fig. 2C-E. Moreover, the EDD dynamics due to the interband dynamics is calculated and plotted in Fig. 4C. In this case, the results show that the interband dynamics is mainly reflected in the change of the EDD localized intensity change around each carbon atom. This change of the EDD occurs alternatively between the carbon atoms. Accordingly, the average intensity change of the EDD is negligible because of the graphene symmetry. Therefore, the interband dynamics doesn't show any observable change of the diffraction peaks intensity or position in our measurements (Fig. 2C-E). Notably, most likely the coherence dynamics would be observed by diffraction measurement in asymmetric systems. Movies of the graphene electron motion related to intraband current and interband dynamics are presented in SM Video 3 & 4, respectively. These results support our conclusion from the fitting of the attosecond diffraction results and explains the direct connection between the light field induced electron dynamics in the CB and the electron motion between the carbon atoms in graphene.

      This work demonstrates the attosecond imaging resolution obtained by attosecond electron pulses inside TEM and the establishment of attomicroscopy. Moreover, we used attomicroscopy to perform attosecond electron diffraction imaging to study the electron dynamics of graphene. Our results built a stronger connection to relate the field-induced electron dynamics in the conduction band and the real time and space electron motion between the carbon atoms in graphene which is not possible by other time-resolved attosecond spectroscopic techniques. Moreover, attomicroscopy imaging provides a connection between light-induced electronic dynamics and the morphology of matter, which builds the desired bridge between science and technology to engineer petahertz and quantum photonics. This work paves the way for recording images of electron motion in the four dimensions of space and time and opens a window to see the quantum world in real systems leading to answering fundamental questions in physics. Furthermore, the demonstrated electron imaging method allows for a vast range of real-life electron imaging applications with potential important impacts in different fields of science and technology. For instance, electron imaging would allow filming and control of chemical and biochemical reactions in real time, which would advance the promising fields of material synthesis, drug design and personalized medicine to the next level.

**References and Notes**

1. F. Krausz, *Electrons in Motion: Attosecond Physics Explores Fastest Dynamics*. (World Scientific Publishing Company Pte. Limited, 2019).
2. P. Corkum, F. Krausz, Attosecond science. *Nat. Phys.* **3**, 381-387 (2007).
3. F. Krausz, M. Ivanov, Attosecond physics. *Rev. Mod. Phys.* **81**, 163 (2009).
4. F. Calegari, D. Ayuso, A. Trabattoni, L. Belshaw, S. De Camillis, S. Anumula, F. Frassetto, L. Poletto, A. Palacios, P. Decleva, J. B. Greenwood, F. Martín, M. Nisoli, Ultrafast electron dynamics in phenylalanine initiated by attosecond pulses. *Science* **346**, 336-339 (2014).
5. G. Sansone, F. Kelkensberg, J. Pérez-Torres, F. Morales, M. F. Kling, W. Siu, O. Ghafur, P. Johnsson, M. Swoboda, E. Benedetti, Electron localization following attosecond molecular photoionization. *Nature* **465**, 763-766 (2010).
6. F. Calegari, G. Sansone, S. Stagira, C. Vozzi, M. Nisoli, Advances in attosecond science. *J Phys. B At. Mol. Opt. Phys.* **49**, 062001 (2016).
7. D. Hui, H. Alqattan, S. Yamada, V. Pervak, K. Yabana, M. T. Hassan, Attosecond electron motion control in dielectric. *Nat. Photon.* **16**, 33-37 (2022).
8. H. Alqattan, D. Hui, M. Sennary, M. T. Hassan, Attosecond electronic delay response in dielectric materials. *Faraday Discuss.*, (2022).
9. B. Wolter, M. G. Pullen, A.-T. Le, M. Baudisch, K. Doblhoff-Dier, A. Senftleben, M. Hemmer, C. D. Schröter, J. Ullrich, T. Pfeifer, R. Moshammer, S. Gräfe, O. Vendrell, C. D. Lin, J. Biegert, Ultrafast electron diffraction imaging of bond breaking in di-ionized acetylene. *Science* **354**, 308-312 (2016).
10. J. Duris, S. Li, T. Driver, E. G. Champenois, J. P. MacArthur, A. A. Lutman, Z. Zhang, P. Rosenberger, J. W. Aldrich, R. Coffee, Tunable isolated attosecond X-ray pulses with gigawatt peak power from a free-electron laser. *Nat. Photon.* **14**, 30-36 (2020).
11. M. Garg, K. Kern, Attosecond coherent manipulation of electrons in tunneling microscopy. *Science* **367**, 411-415. (2020).
12. A. H. Zewail, 4D ultrafast electron diffraction, crystallography, and microscopy. *Annu. Rev. Phys. Chem.* **57**, 65-103 (2006).
13. K. Amini, J. Biegert, in *Advances In Atomic, Molecular, and Optical Physics,* L. F. Dimauro, H. Perrin, S. F. Yelin, Eds. (Academic Press, 2020), vol. 69, pp. 163-231.
14. J. Yang, X. Zhu, J. P. F. Nunes, K. Y. Jimmy, R. M. Parrish, T. J. Wolf, M. Centurion, M. Gühr, R. Li, Y. Liu, Simultaneous observation of nuclear and electronic dynamics by ultrafast electron diffraction. *Science* **368**, 885-889 (2020).
15. R. J. D. Miller, Femtosecond Crystallography with Ultrabright Electrons and X-rays: Capturing Chemistry in Action. *Science* **343**, 1108-1116 (2014).
16. A. H. Zewail, Four-dimensional electron microscopy. *Science* **328**, 187-193 (2010).
17. L. Ma, H. Yong, J. D. Geiser, A. Moreno Carrascosa, N. Goff, P. M. Weber, Ultrafast x-ray and electron scattering of free molecules: A comparative evaluation. *Struct. Dyn.* **7**, 034102 (2020).
18. M. T. Hassan, J. S. Baskin, LiaoB, A. H. Zewail, High-temporal-resolution electron microscopy for imaging ultrafast electron dynamics. *Nat. Photon.* **11**, 425-430 (2017).
19. M. T. Hassan, Attomicroscopy: from femtosecond to attosecond electron microscopy. *J Phys. B At. Mol. Opt. Phys.* **51**, 032005 (2018).
20. B. E. Warren, *X-ray Diffraction*. (Courier Corporation, 1990).

21. H. Ihee, V. A. Lobastov, U. M. Gomez, B. M. Goodson, R. Srinivasan, C.-Y. Ruan, A. H. Zewail, Direct imaging of transient molecular structures with ultrafast diffraction. *Science* **291**, 458-462 (2001).
22. B. J. Siwick, J. R. Dwyer, R. E. Jordan, R. D. Miller, An atomic-level view of melting using femtosecond electron diffraction. *Science* **302**, 1382-1385 (2003).
23. R. Ernstorfer, M. Harb, C. T. Hebeisen, G. Sciaini, T. Dartigalongue, R. D. Miller, The formation of warm dense matter: experimental evidence for electronic bond hardening in gold. *Science* **323**, 1033-1037 (2009).
24. R. P. Chatelain, V. R. Morrison, C. Godbout, B. J. Siwick, Ultrafast electron diffraction with radio-frequency compressed electron pulses. *Appl. Phys. Lett.* **101**, 081901 (2012).
25. M. Gao, C. Lu, H. Jean-Ruel, L. C. Liu, A. Marx, K. Onda, S.-y. Koshihara, Y. Nakano, X. Shao, T. Hiramatsu, Mapping molecular motions leading to charge delocalization with ultrabright electrons. *Nature* **496**, 343-346 (2013).
26. T. Van Oudheusden, P. Pasmans, S. Van Der Geer, M. De Loos, M. Van Der Wiel, O. Luiten, Compression of subrelativistic space-charge-dominated electron bunches for single-shot femtosecond electron diffraction. *Phys. Rev. Lett.* **105**, 264801 (2010).
27. V. R. Morrison, R. P. Chatelain, K. L. Tiwari, A. Hendaoui, A. Bruhács, M. Chaker, B. J. Siwick, A photoinduced metal-like phase of monoclinic VO2 revealed by ultrafast electron diffraction. *Science* **346**, 445-448 (2014).
28. A. Feist, K. E. Echternkamp, J. Schauss, S. V. Yalunin, S. Schäfer, C. Ropers, Quantum coherent optical phase modulation in an ultrafast transmission electron microscope. *Nature* **521**, 200-203 (2015).
29. J. Yang, X. Zhu, T. J. Wolf, Z. Li, J. P. F. Nunes, R. Coffee, J. P. Cryan, M. Gühr, K. Hegazy, T. F. Heinz, Imaging CF3I conical intersection and photodissociation dynamics with ultrafast electron diffraction. *Science* **361**, 64-67 (2018).
30. M. R. Otto, L. P. René de Cotret, M. J. Stern, B. J. Siwick, Solving the jitter problem in microwave compressed ultrafast electron diffraction instruments: Robust sub-50 fs cavity-laser phase stabilization. *Struct. Dyn.* **4**, 051101 (2017).
31. K. E. Priebe, C. Rathje, S. V. Yalunin, T. Hohage, A. Feist, S. Schäfer, C. Ropers, Attosecond electron pulse trains and quantum state reconstruction in ultrafast transmission electron microscopy. *Nat. Photon.* **11**, 793-797 (2017).
32. Y. Morimoto, P. Baum, Diffraction and microscopy with attosecond electron pulse trains. *Nat. Phys.* **14**, 252-256 (2018).
33. A. Ryabov, J. W. Thurner, D. Nabben, M. V. Tsarev, P. Baum, Attosecond metrology in a continuous-beam transmission electron microscope. *Science advances* **6**, eabb1393 (2020).
34. H. W. Kim, N. A. Vinokurov, I. H. Baek, K. Y. Oang, M. H. Kim, Y. C. Kim, K.-H. Jang, K. Lee, S. H. Park, S. Park, Towards jitter-free ultrafast electron diffraction technology. *Nat. Photon.* **14**, 245-249 (2020).
35. M. Kozák, J. McNeur, K. Leedle, H. Deng, N. Schönenberger, A. Ruehl, I. Hartl, J. Harris, R. Byer, P. Hommelhoff, Optical gating and streaking of free electrons with sub-optical cycle precision. *Nat. Commun.* **8**, 14342 (2017).
36. C. M. S. Sears, E. Colby, R. Ischebeck, C. McGuinness, J. Nelson, R. Noble, R. H. Siemann, J. Spencer, D. Walz, T. Plettner, R. L. Byer, Production and characterization of attosecond electron bunch trains. *Phys. Rev. Accel. Beams* **11**, 061301 (2008).
37. M. T. Hassan, H. Liu, J. S. Baskin, A. H. Zewail, Photon gating in four-dimensional ultrafast electron microscopy. *Proc. Natl. Acad. Sci. U.S.A.* **112**, 12944-12949 (2015).

38. G. Sansone, E. Benedetti, F. Calegari, C. Vozzi, L. Avaldi, R. Flammini, L. Poletto, P. Villoresi, C. Altucci, R. Velotta, S. Stagira, S. De Silvestri, M. Nisoli, Isolated Single-Cycle Attosecond Pulses. *Science* **314**, 443 (2006).
39. M. Nisoli, G. Sansone, New frontiers in attosecond science. *Prog. Quantum. Electron.* **33**, 17-59 (2009).
40. M. I. Stockman, Nanoplasmonics: past, present, and glimpse into future. *Opt. Express* **19**, 22029-22106 (2011).
41. G. M. Vanacore, G. Berruto, I. Madan, E. Pomarico, P. Biagioni, R. Lamb, D. McGrouther, O. Reinhardt, I. Kaminer, B. Barwick, Ultrafast generation and control of an electron vortex beam via chiral plasmonic near fields. *Nature materials* **18**, 573-579 (2019).
42. T. Higuchi, C. Heide, K. Ullmann, H. B. Weber, P. Hommelhoff, Light-field-driven currents in graphene. *Nature* **550**, 224-228 (2017).
43. T. Boolakee, C. Heide, A. Garzón-Ramírez, H. B. Weber, I. Franco, P. Hommelhoff, Light-field control of real and virtual charge carriers. *Nature* **605**, 251-255 (2022).
44. M. Baudisch, A. Marini, J. D. Cox, T. Zhu, F. Silva, S. Teichmann, M. Massicotte, F. Koppens, L. S. Levitov, F. J. García de Abajo, Ultrafast nonlinear optical response of Dirac fermions in graphene. *Nat. Commun.* **9**, 1-6 (2018).
45. E. McCann, M. Koshino, The electronic properties of bilayer graphene. *Rep. Prog. Phys.* **76**, 056503 (2013).
46. J. Krieger, G. Iafrate, Time evolution of Bloch electrons in a homogeneous electric field. *Physical Review B* **33**, 5494 (1986).
47. C. Liu, Y. Zheng, Z. Zeng, R. Li, Driving-laser ellipticity dependence of high-order harmonic generation in graphene. *Phys. Rev. A* **97**, 063412 (2018).
48. H. Alqattan, D. Hui, V. Pervak, M. T. Hassan, Attosecond light field synthesis. *APL Photonics* **7**, 041301 (2022).

**Acknowledgments:**

This project is funded in part by Gordon and Betty Moore Foundation Grant GBMF7938 to M. Hassan. Additionally, this material is based upon work partially supported by the Air Force Office of Scientific Research under award number FA9550-19-1-0025. We are also grateful to the W.M. Keck Foundation for supporting this project with a Science and Engineering award given to M.Th.H. Moreover, N.V.G. acknowledges support from the Branco Weiss Fellowship—Society in Science, administered by ETH Zürich.

**Author contributions:**

D.H., H.A. and M.Th.H. built the Attosecond Electron Microscopy experimental setup. D.H., H.A., M.S., and M.Th.H. conducted the experiments and analysed the data. N.V.G. developed the theoretical model and performed the numerical simulations. M.Th.H. conceived, supervised, and directed the study. All authors discussed the results and their interpretation and wrote the manuscript.

# Figures and Figure captions

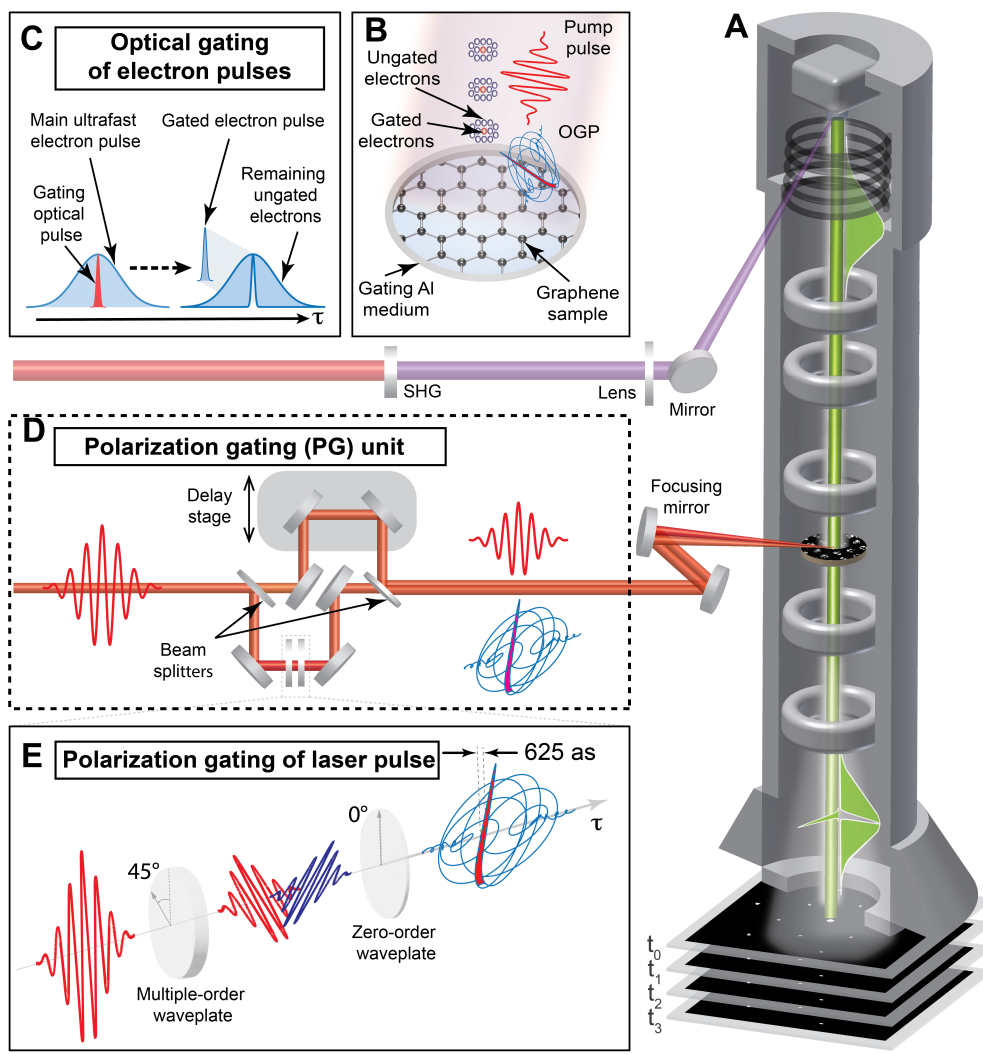

**Fig. 1. Attomicroscopy and the generation of attosecond electron pulse**. **A**, Attomicroscopy setup. **B**, Illustration of the electron and laser pulse interaction on the gating medium (aluminum grid) above the sample. **C**, The basic principle of the optical gating of ultrafast electron pulse by optical laser pulse. The gating time window and the duration of the generated gated electrons are similar to the laser pulse duration. **D**, Schematic of the polarization gating (PG) setup. In the PG, a 5-fs laser pulse is divided by a beamsplitter into two beams. The first beam is reflected off two mirrors mounted on a nanometer-precision delay stage and is sent to the microscope. This beam is used as the pump pulse to trigger the electron motion in the system under study. The second beam undergoes the polarization gating process, which is illustrated in e. In this process, the input linearly polarized laser pulse pass through multi-order and zero-order waveplates set at 45 and 0 degrees, respectively. The generated pulse has a half-cycle linear-polarized field in the middle and a circularly polarized field at the edges. This pulse is used as optical gating pulse (OGP) of the electrons inside the microscope. The OGP beam is directed to the sample position where the electron gating and the generation of the attosecond electron pulses (AEPs) takes place. The AEPs are used to probe the triggered electron motion in the sample with a pump pulse. A series of images (direct/diffraction) can be recorded to obtain a movie of electron motion in action.

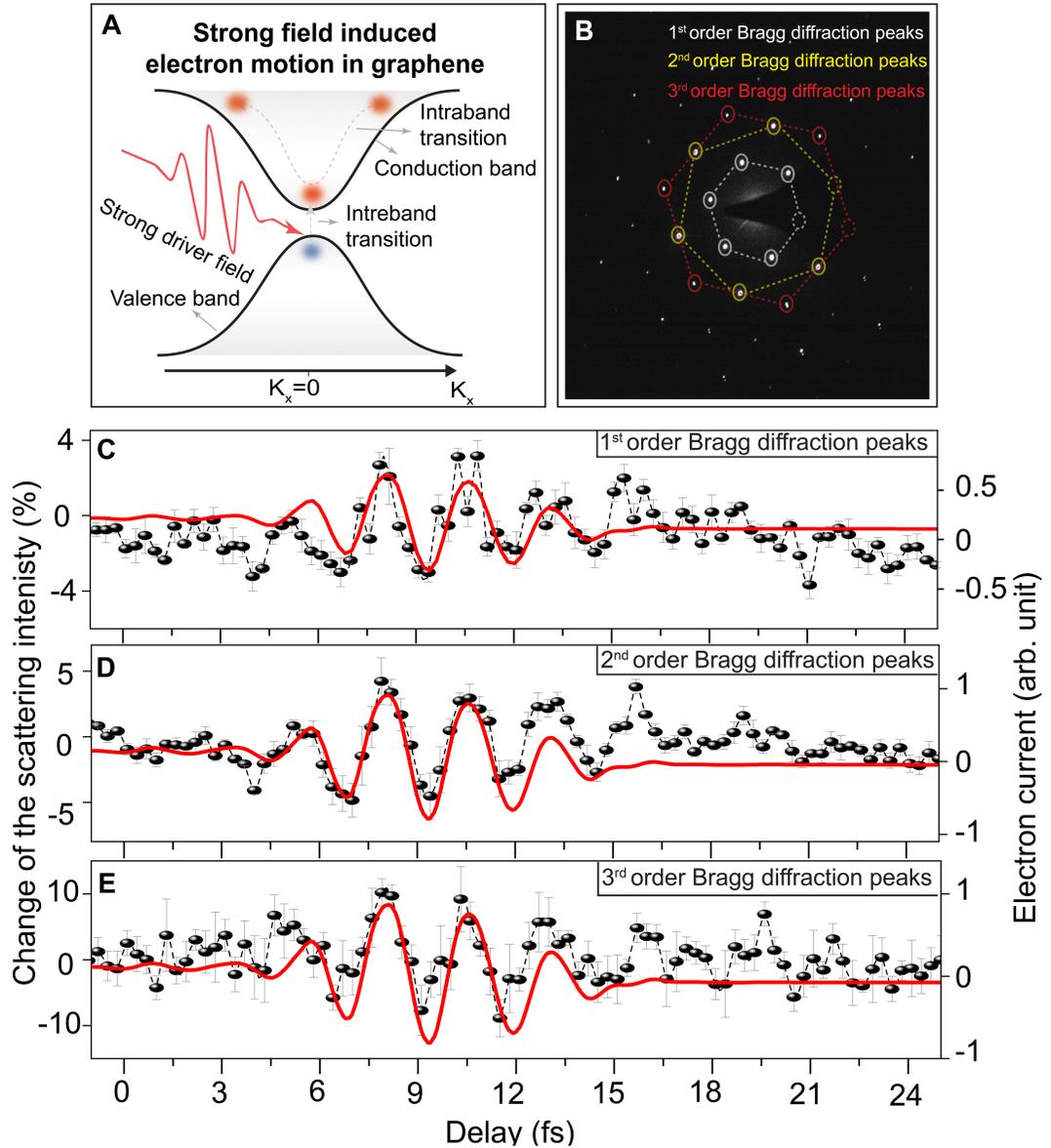

**Fig. 2. Attosecond electron diffraction imaging of electron motion dynamics in graphene. A**, Illustration of the strong light field-driven electron motion mechanism in graphene. **B**, Electron diffraction pattern of the multilayer graphene sample. The white, yellow, and red dashed lines indicate the 1st-, 2nd-, and 3rd-order diffraction peaks, respectively. **C-E**, The retrieved average scattering intensity changes of the 1$^{st}$-, 2$^{nd}$-, and 3$^{rd}$-order diffraction peaks as they evolve over time are shown as black dots connected with dashed black lines. The error bars present standard deviation uncertainty of the relative scattered peak intensity of 7 scans. The fitted calculated field-induced electronic currents in graphene are plotted in solid red line. The fitting line show good agreement with the measured diffraction intensity oscillations.

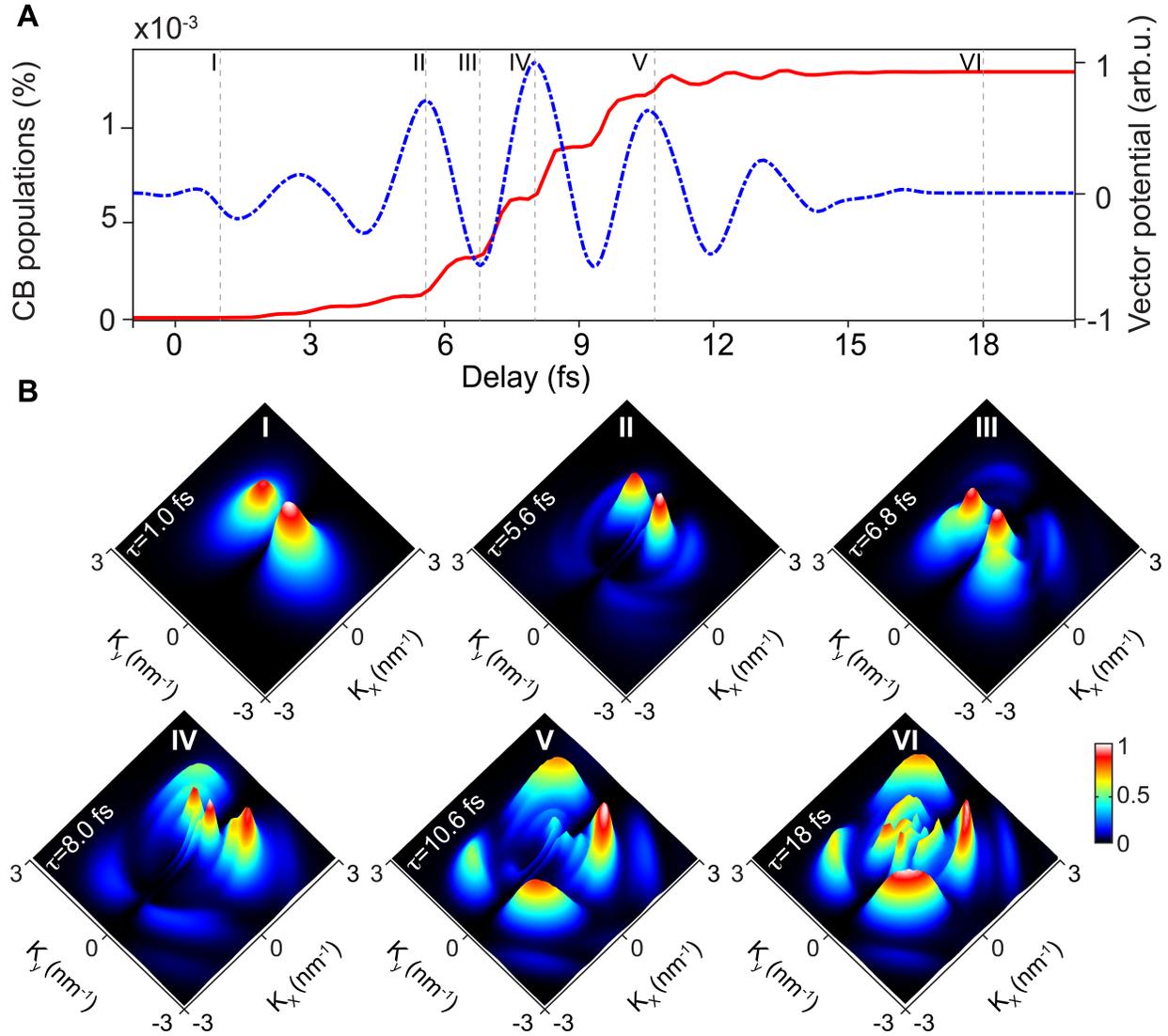

**Fig. 3. Light field-induced electron motion dynamics in the reciprocal space of graphene**. **A**, The calculated total electron population dynamics in the conduction band (red solid line) is plotted in contrast with the vector potential of the driver field (blue dashed line). **B(I-VI)**, Snapshots of the electron density distribution at different time instants (indicated by dashed grey lines in A) in the presence of and after the end of the driver field. Each snapshot is normalized respect to the snapshot in BVI for better visualizing of the electron density distribution (the normalization factor is mentioned in the right side inset).

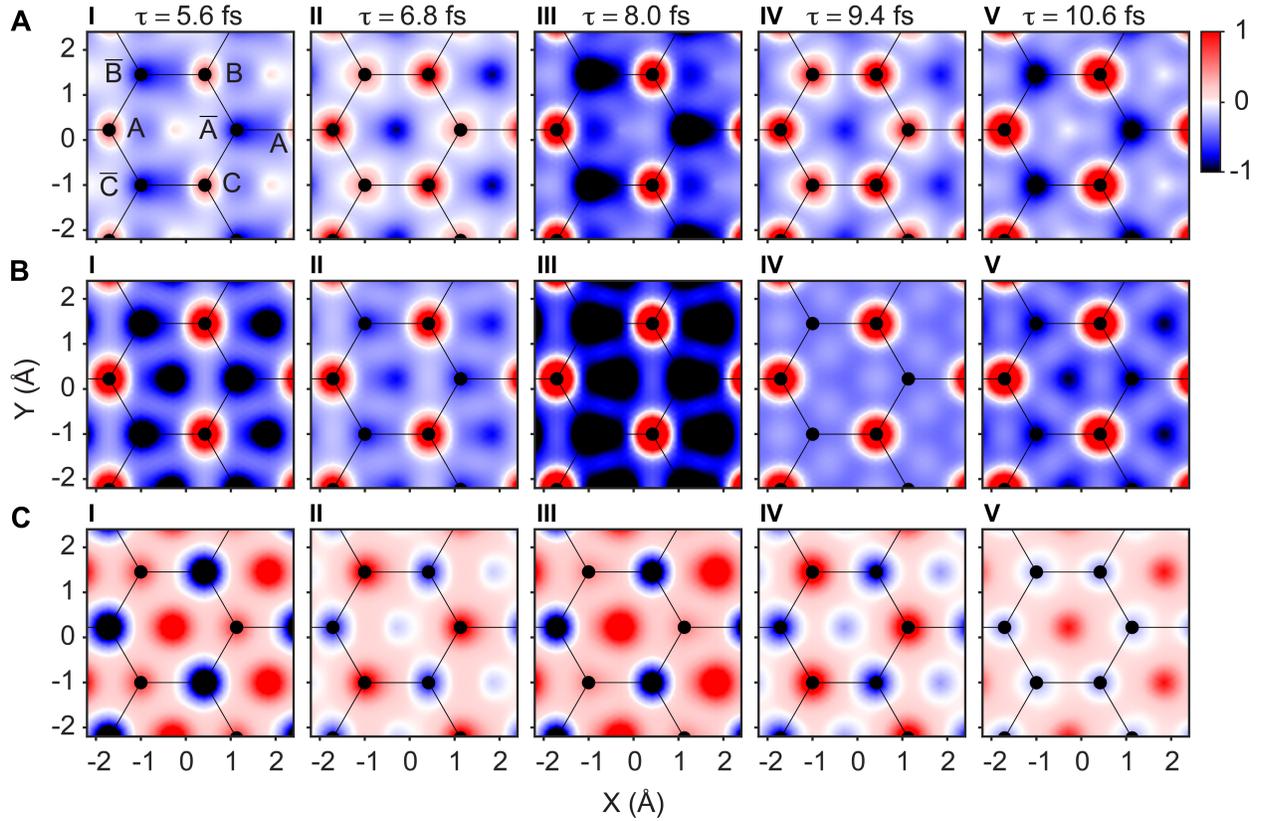

**Fig. 4. Graphene electron motion dynamics in real time and space**. **A(I-V)**, Snapshots of the field-induced electron motion at different time instants. These snapshots show the change in the electron density distribution (EDD)—after subtracting the density in the equilibrium state before the field arrival—around and between the carbon atoms of graphene at different time instants in real space. The red, blue, and white colors represent the high (positive), the low (negative) and zero values of the EDD, respectively. **B(I-V)**, The corresponding snapshoots showing the EDD dynamics solely due to the intraband dynamics. The change of the EDD is mainly occurring between the c-c atoms. **C(I-V)**, The real space EDD dynamics due to the interband dynamics. The EDD intensity changes locally around the carbon atoms in an alternative manner.